\documentclass[onecolumn,floatfix,aps,superscriptaddress]{revtex4}
\usepackage{graphicx}
\usepackage{amsmath}
\usepackage[colorlinks=true, citecolor=blue, urlcolor=blue, linkcolor = blue]{hyperref}
\usepackage{amssymb}
\usepackage{bm}
\usepackage{color}
\usepackage{bookmark}
\usepackage{tabularx}
\usepackage{mathtools}
\usepackage{microtype}
\usepackage{cancel}
\usepackage{relsize}

\usepackage{xcolor}
\begin{document}
\title{Charged space debris induced nonlinear magnetosonic waves using inertial magnetohydrodynamics }
\author{S. P. Acharya
\footnote{Electronic mail: siba.acharya@saha.ac.in and siba.acharya39@gmail.com}}
\affiliation{Saha Institute of Nuclear Physics, a Constituent Institute of Homi Bhabha National Institute (HBNI), 1/AF Bidhannagar, Kolkata-700064 (India)}
\author{A. Mukherjee
\footnote{Electronic mail: abhikmukherjeesinp15@gmail.com}}
\affiliation{Physics and Applied Mathematics Unit, Indian Statistical Institute, Kolkata, (India)}
\author{M. S. Janaki
\footnote{Electronic mail: ms.janaki@saha.ac.in}}
\affiliation{Saha Institute of Nuclear Physics, a Constituent Institute of Homi Bhabha National Institute (HBNI), 1/AF Bidhannagar, Kolkata-700064 (India)}
\begin{abstract}
The excitations of nonlinear magnetosonic waves in presence of charged space debris in the low Earth orbital plasma region is investigated taking into account effects of electron inertia in the framework of classical magnetohydrodynamics, which is also referred to as inertial magnetohydrodynamics. Magnetosonic waves are found to be governed by a  forced Kadomtsev-Petviashvili equation with the forcing term representing effects of space debris particles. The dynamical behaviours of both slow and fast magnetosonic solitary waves is explored in detail. Exact accelerated magnetosonic lump solutions are shown to be stable for the entire region in parameter space of slow waves and a large region in parameter space of fast waves. In a similar way, magnetosonic curved solitary waves become stable for a small region in parameter space of fast waves. These exact solutions with special properties are derived for specific choices of debris functions. These novel results can have potential applications in scientific and technological aspects of space debris detection and mitigation.
\end{abstract}
\maketitle
\section{Introduction}
There have been many scientific and technological advancements in the dynamic research field dealing with space plasmas in recent years. In this context, near-Earth space plasma environment has gained significant attention across different countries in the world as it is extremely important for different space missions. This has turned this plasma system into a virtual vast laboratory for studying different scientific and technological phenomena through measurements including both in situ and ground-based.

 There have been reported many theoretical, computational and equivalent experimental investigations of space debris dynamics in near-Earth space plasma environment \cite{Kulikov, Sen, Sen3, Mukherjee, Acharya, Truitt, TruittKP, Kumar, Acharyaa} done in recent years. The motion of charged space debris particles and their interaction with the surrounding plasma in ionosphere can give rise to various types of nonlinear structures such as precursors and solitons with electrostatic
and electromagnetic  characteristics. In the context of electrostatic
excitations, the propagation of weakly nonlinear dispersive ion acoustic waves driven by space debris particles has been modelled in the
form of a forced Korteweg-de Vries (fKdV) equation \cite{Sen} or forced Kadomtsev-Petviashvili (fKP) equation \cite{TruittKP, Acharya}. For a moving charge bunch in a magnetized plasma, particle-in-cell
simulations by Kumar and Sen \cite{Kumar} have established for the first time the existence of precursor and pinned magnetosonic solitons; which are also applicable for charged space debris motion in the ionospheric plasma. The importance of such electromagnetic structures lies in their easy detection from remote distances due to their spatial extent being a few electron skin depths; thereby permitting larger footprints. In the framework of Hall magnetohydrodynamics (MHD) \cite{Ruderman} model, theoretical studies of electromagnetic nonlinear structures excited by moving charged debris particles modelled as moving current sources have lead to lump wave
structures for sources having two dimensional nature as reported in our recent work \cite{Acharyaa}. It should be noted that only the term representing Hall current in Ohm's law  is retained in case of Hall MHD. Hence, by considering Hall MHD in our previous work \cite{Acharyaa}, we have neglected electron inertial effects, and, as a consequence, dispersive effect vanishes for perpendicular propagation with respect to ambient magnetic field direction that can be seen in \cite{Acharyaa}. Therefore, for describing space debris dynamics in the Low Earth Orbital (LEO) \cite{Sampaio} or ionospheric plasma region from a general viewpoint along with finite dispersive effects, we need to invoke inertial effects of electrons by considering their masses in full MHD model. In the present
work, a study of nonlinear magnetosonic solitary wave excitations by current 
sources is proposed by considering the classical MHD model with inclusion of
finite electron inertial effects, i.e. an extended MHD or inertial MHD model \cite{Lingam}. The inclusion of inertial effects further
enables study of both oblique and perpendicular propagations with respect to magnetic field due to non-zero dispersive effects at all angles of propagation. The proposed inertial MHD model in this article to investigate space debris dynamics in the LEO region can also enact as a generalization to our previous work \cite{Acharyaa}. As pointed out earlier, this model is particularly more significant for perpendicular propagation than propagations at oblique angles with respect to ambient magnetic field direction in the LEO plasma; which is the crucial finding of this article as explored in the latter part in detail.
 

This article is organized in the following manner. The detailed derivation of (2+1) dimensional nonlinear evolution equation in the form of forced KP equation is given in section- II. The dynamics of magnetosonic solitary waves arising due to presence of charged debris field is discussed in section- III. Conclusive remarks are provided in section -IV followed by acknowledgements and bibliography.

\section{Derivation of the nonlinear evolution equation for magnetosonic waves in presence of charged debris using inertial magnetohydrodynamics}
We consider the dynamical evolution of nonlinear magnetosonic waves in the LEO plasma region. This is a low temperature and low density plasma in presence of orbital motions of charged space debris particles, which are ubiquitous in near-Earth space. The ambient magnetic field in this plasma region arising due to the Earth's magnetosphere, interplanetary magnetic fields etc. is also taken into account, which plays a vital role in propagation of nonlinear magnetosonic waves in our system. The dynamics of these hydromagnetic waves can suitably be described by equations of inertial magnetohydrodynamics (MHD), that are given as: 
\begin{equation}
\frac{\partial n}{\partial t}+\vec{\nabla}.(n\vec{v})=0, \label{Continuity}
\end{equation}
\begin{equation}
\frac{d \vec{v}}{d t}=\mu \frac{d}{d t}\frac{\vec{\nabla}\times\vec{B}}{n}+\frac{(\vec{\nabla}\times\vec{B})\times\vec{B}}{n}-{\beta}\frac{\vec{\nabla}n}{n}+\mu (\frac{\vec{\nabla}\times\vec{B}}{n}.\vec{\nabla})\vec{v}-\mu(\frac{\vec{\nabla}\times\vec{B}}{n}.\vec{\nabla})(\frac{\vec{\nabla}\times\vec{B}}{n})-\frac{\vec{J_s}\times\vec{B}}{n}, \label{Momentum}
\end{equation}
\begin{equation}
\frac{\partial \vec{B}}{\partial t}=\vec{\nabla}\times(\vec{v}\times\vec{B})-\vec{\nabla}\times\frac{d \vec{v}}{d t}, \label{OhmLaw}
\end{equation}
where we have used the work of Janaki et al. \cite{Janaki} in writing our model equations. The equations (\ref{Continuity}), (\ref{Momentum}) and (\ref{OhmLaw}) represent continuity equation, momentum equation and Ohm's law respectively. Here $n$ is ion number density, $\vec{v}$ is (3D) fluid velocity, $\vec{B}$ is ambient magnetic field in the LEO region and $\vec{J}_s$ is source current due to space debris particles. $\beta$ and $\mu$ are constants given by
\begin{equation}
\beta =\frac{c^2_s}{v^2_A}; \mu =\frac{m_e}{m_i} \label{beta}
\end{equation}
where $c_s$ and $v_A$ denote sound speed and Alfven speed respectively. The constant $\mu$ stands for inertial effects of electrons and ions which was neglected in our earlier work \cite{Acharyaa}. It should be emphasized here that in the limit $\mu \longrightarrow 0$, our model equations given by equations (\ref{Continuity}), (\ref{Momentum}) and (\ref{OhmLaw}) converge to Hall MHD equations taken in \cite{Acharyaa}. The above equation (\ref{beta}) also implies that value of $\beta$ for a given system can be determined from values of ion density, ion temperature and strength of ambient magnetic field in the system. In writing the above model equations (\ref{Continuity}), (\ref{Momentum}) and (\ref{OhmLaw}), following normalizations have been used:
\begin{equation}
n \longrightarrow \frac{n}{n_0};\,\vec{v} \longrightarrow \frac{\vec{v}}{v_A};\,\vec{B} \longrightarrow \frac{\vec{B}}{B_0};\,t \longrightarrow {\Omega}_i t;\, \vec{\nabla} \longrightarrow \frac{v_A}{{\Omega}_i }\vec{\nabla} ,\label{NormMHD}
\end{equation}
where $n_0,$  $B_0$ denote equilibrium ion number density and equilibrium (ambient) magnetic field respectively, and ${\Omega}_i $ stands for ion cyclotron frequency. 
We assume that nonlinear waves have weak transverse propagation. We consider that the source debris current resulting the motion of debris in magnetized plasma medium of the LEO region is approximated to be in $z$ direction. In order to derive the appropriate nonlinear evolution equation (NLEE) governing the dynamics of nonlinear magnetosonic waves using well-known reductive perturbation technique (RPT), we expand the dependent variables of our system as:
\begin{equation}
n=1+\epsilon {n}_1+{\epsilon}^2 {n}_2+O({\epsilon}^3),\label{RhoExpn}
\end{equation}
\begin{equation}
v_x=\epsilon v_{x1}+{\epsilon}^2 v_{x2}+O({\epsilon}^3);\,
v_y=\epsilon v_{y1}+{\epsilon}^2 v_{y2}+O({\epsilon}^3);\,
v_z={\epsilon}^{3/2} v_{z1}+{\epsilon}^{5/2} v_{z2}+O({\epsilon}^{7/2}),\label{VExpn}
\end{equation}
\begin{equation}
B_x=cos (\theta);\,
B_y=sin (\theta)+\epsilon B_{y1}+{\epsilon}^2 B_{y2}+O({\epsilon}^3);\,
B_z={\epsilon}^{3/2} B_{z1}+{\epsilon}^{5/2} B_{z2}+O({\epsilon}^{7/2}),\label{BExpn}
\end{equation}
\begin{equation}
J_{sz}={\epsilon}^{5/2}J_D, \label{JExpn}
\end{equation}
where $\epsilon$ is a small dimensionless parameter characterizing the strength of nonlinearity, and $J_D$ is the new vriable related to source debris current in the system. Here $\theta$ is the angle which magnetic field $\vec{B}$ makes with $x-$axis. For simplicity, it is assumed that dependent variables do not change along $y$ direction. This implies $\vec{\nabla}=(\frac{\partial}{\partial x},0,\frac{\partial}{\partial z})$. Since we have considered weak transverse propagation, scalings of $v_z$ and $J_{sz}$ are weaker as compared to other components.

The stretched variables for weak transverse perturbations are introduced as
\begin{equation}
\xi={\epsilon}^{1/2}(x-v_p t);\,\tau ={\epsilon}^{3/2} t;\,\zeta=\epsilon z. \label{IndptScal}
\end{equation}

\begin{figure}[hbt!]
\centering
\includegraphics[width=17cm]{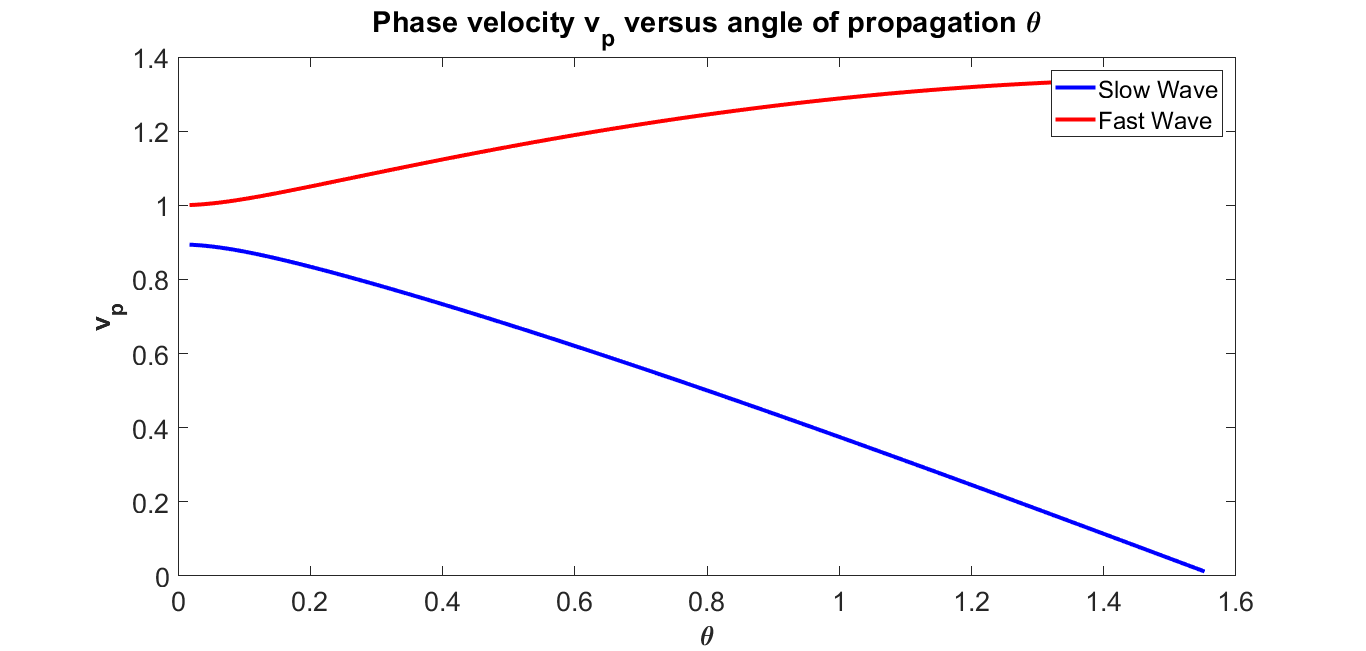}
\caption{Variation of phase velocity  $v_p$ given by equation (\ref{DispRela}) against angle of propagation $\theta$ for both slow and fast waves considering $\beta=0.8$, with $\theta$ expressed in radian unit}
\label{Plota}
\end{figure}

Now following the rigorous calculations order by order using reductive 
perturbation theory, we get the following relations at the order $O({\epsilon}^{3/2})$:
\begin{equation}
v_{x1}=v_pn_1;\,v_{y1}=\frac{{\beta}-v^2_p}{v_p}cot(\theta)n_1;\,B_{y1}=\frac{v^2_p-{\beta}}{sin(\theta)}n_1, \label{VxyBy1}
\end{equation}
which is accompanied with the dispersion relation given by
\begin{equation}
v^4_p-(1+{\beta})v^2_p+{\beta}cos^2(\theta)=0\label{DispRel}.
\end{equation}
This implies that the value of phase velocity $v_p$ is calculated as:
\begin{equation}
v^2_p=\frac{(1+{\beta})\pm\sqrt{{(1+{\beta})}^2-4{\beta}cos^2(\theta)}}{2}. \label{DispRela}
\end{equation}
The dispersion relation (\ref{DispRel}) clearly indicates that the nonlinear waves, which we are dealing with, are of magnetosonic types with slow and fast modes. The typical variation of phase velocity given by above equation (\ref{DispRela}) against angle of propagation $\theta$ for both slow and fast waves is shown in Figure-\ref{Plota}.

In a similar manner, from equations of the order $O({\epsilon}^{5/2})$, we get
$$
[v_p(1-v^2_p)][\frac{\partial n_1}{\partial \tau}+\frac{\partial}{\partial \xi}(n_1 v_{x1})+\frac{\partial v_{z1}}{\partial \zeta}]+[cos^2(\theta)-v^2_p][\frac{\partial v_{x1}}{\partial \tau}+v_{x1}\frac{\partial v_{x1}}{\partial \xi}-\mu v_p\frac{{\partial}^2B_{y1}}{\partial \xi \partial \zeta}+B_{y1}\frac{\partial B_{y1}}{\partial \xi}-sin(\theta)n_1\frac{B_{y1}}{\partial \xi}$$
$$
-\beta n_1\frac{\partial n_1}{\partial \xi}-J_D sin(\theta)]+[sin(\theta)cos(\theta)][\frac{\partial v_{y1}}{\partial \tau}+v_{x1}\frac{\partial v_{y1}}{\partial \xi}- \mu v_p\frac{{\partial}^2B_{z1}}{\partial {\xi}^2}+cos(\theta)n_1\frac{\partial B_{y1}}{\partial \xi}+J_D cos(\theta)]-[v_psin(\theta)][\frac{\partial B_{y1}}{\partial \tau}
$$
\begin{equation}
+sin(\theta)\frac{\partial v_{z1}}{\partial \zeta}+\frac{\partial}{\partial \xi}(v_{x1}B_{y1})-v_p\frac{{\partial}^2v_{x1}}{\partial \zeta \partial \xi} +v_p\frac{{\partial}^2v_{z1}}{\partial {\xi}^2}]=0. \label{ABCDEqn}
\end{equation} 
Differentiating the above equation (\ref{ABCDEqn}) with respect to $\xi$  and using equations of order $O({\epsilon}^2)$ and equation (\ref{VxyBy1}), we get the following equation: 
$$
\frac{\partial}{\partial \xi}\lbrace \frac{\partial n_1}{\partial \tau}+\frac{v^3_p(2-5v^2_p+3\beta)}{v^2_p(1+\beta-3v^2_p)+\beta cos^2(\theta)}n_1\frac{\partial n_1}{\partial \xi}+\frac{v^3_p(v^2_p-\beta)[(\mu-{\mu}^2-1)cos^2(\theta)+\mu v^2_p]}{[cos^2(\theta)-v^2_p][v^2_p(1+\beta-3v^2_p)+\beta cos^2(\theta)]}\frac{{\partial}^3n_1}{\partial {\xi}^3} \rbrace$$
\begin{equation}
+\frac{v^5_p}
{v^2_p(3v^2_p-\beta-1)-\beta cos^2(\theta)}\frac{{\partial}^2n_1}{\partial {\zeta}^2}=\frac{v^3_p sin(\theta)}{v^2_p(3v^2_p-\beta-1)-\beta cos^2(\theta)}\frac{\partial J_D}{\partial \xi}, \label{FinalKP1}
\end{equation}
which is the final nonlinear evolution equation in the form of forced Kadomtsev-Petviashvili (fKP) equation describing the dynamics of nonlinear magnetosonic waves in our system. In compact form, this fKP equation (\ref{FinalKP1}) can be written as
\begin{equation}
\frac{\partial}{\partial \xi}(\frac{\partial n_1}{\partial \tau}+N n_1\frac{\partial n_1}{\partial \xi}+D\frac{{\partial}^3 n_1}{\partial {\xi}^3})+T\frac{{\partial}^2 n_1}{\partial {\zeta}^2}=E\frac{\partial J_D}{\partial {\xi}}, \label{FinalKP}
\end{equation}
where $N,\,D,\,T$ and $E$ denote coefficients corresponding to the nonlinear, dispersive, $2D$ and external forcing terms in fKP equation (\ref{FinalKP1}) respectively, and are given by
$$N=\frac{v^3_p(2-5v^2_p+3\beta)}{v^2_p(1+\beta-3v^2_p)+\beta cos^2(\theta)};\,D=\frac{v^3_p(v^2_p-\beta)[(\mu-{\mu}^2-1)cos^2(\theta)+\mu v^2_p]}{[cos^2(\theta)-v^2_p][v^2_p(1+\beta-3v^2_p)+\beta cos^2(\theta)]};$$
\begin{equation}
T=\frac{v^5_p}{v^2_p(3v^2_p-\beta-1)-\beta cos^2(\theta)};\,E=\frac{v^3_p sin(\theta)}{v^2_p(3v^2_p-\beta-1)-\beta cos^2(\theta)}. \label{NDTE}
\end{equation}
In the limit $\mu \longrightarrow 0$, the above coefficients $N,\,D,\,T$ and $E$ become identical to those derived in our previous work on Hall plasma \cite{Acharyaa} as well. The important point is that the dispersive term $D$ vanishes for perpendicular propagation with respect to ambient magnetic field direction when $\mu \longrightarrow 0$. This implies that if we neglect electron inertial effects by considering zero electron mass, then dispersive effect vanishes for propagations of nonlinear waves in the direction perpendicular to ambient magnetic field. This characteristic is also evident in \cite{Acharyaa} in the context of Hall MHD formulation of space debris problem where electron inertial effects are neglected. In the following section, we will analyze different interesting solutions of fKP equation (\ref{FinalKP}) to explain the wave dynamics in case of both slow and fast magnetosonic waves.
\section{Dynamics of nonlinear magnetosonic waves in presence of charged space debris}
\begin{figure}[hbt!]
\centering
\includegraphics[width=17cm]{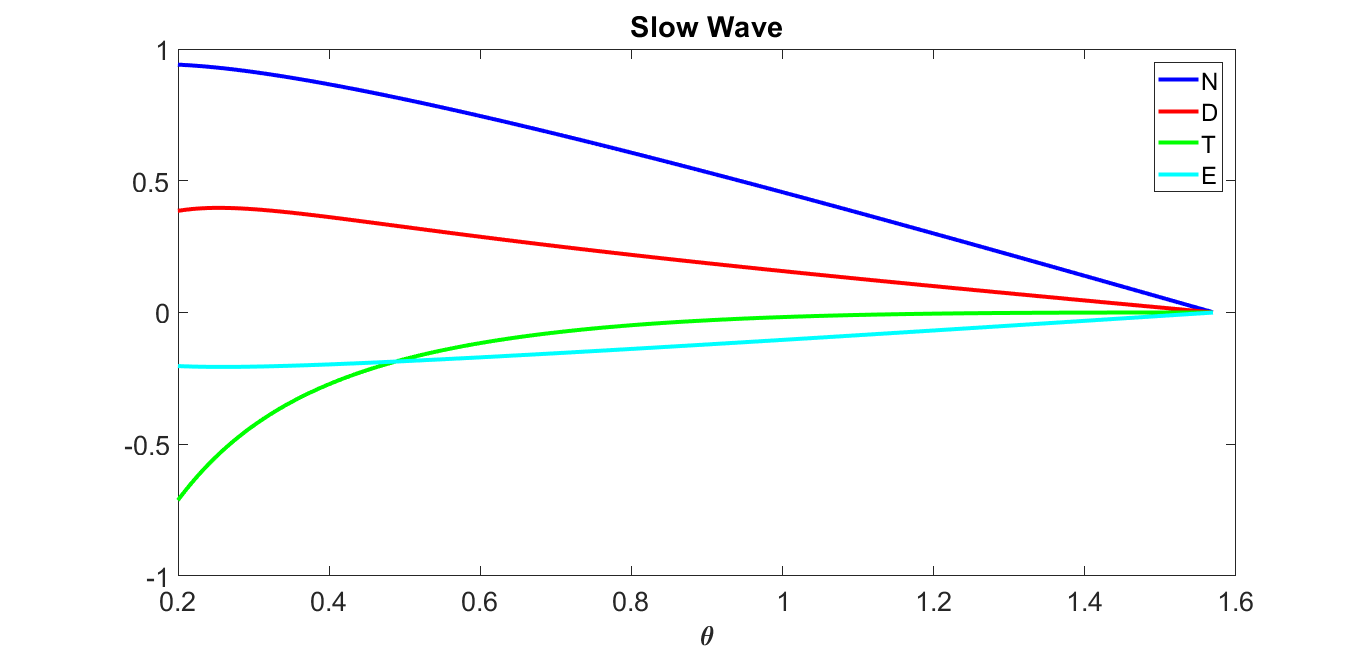}
\ \
\includegraphics[width=17cm]{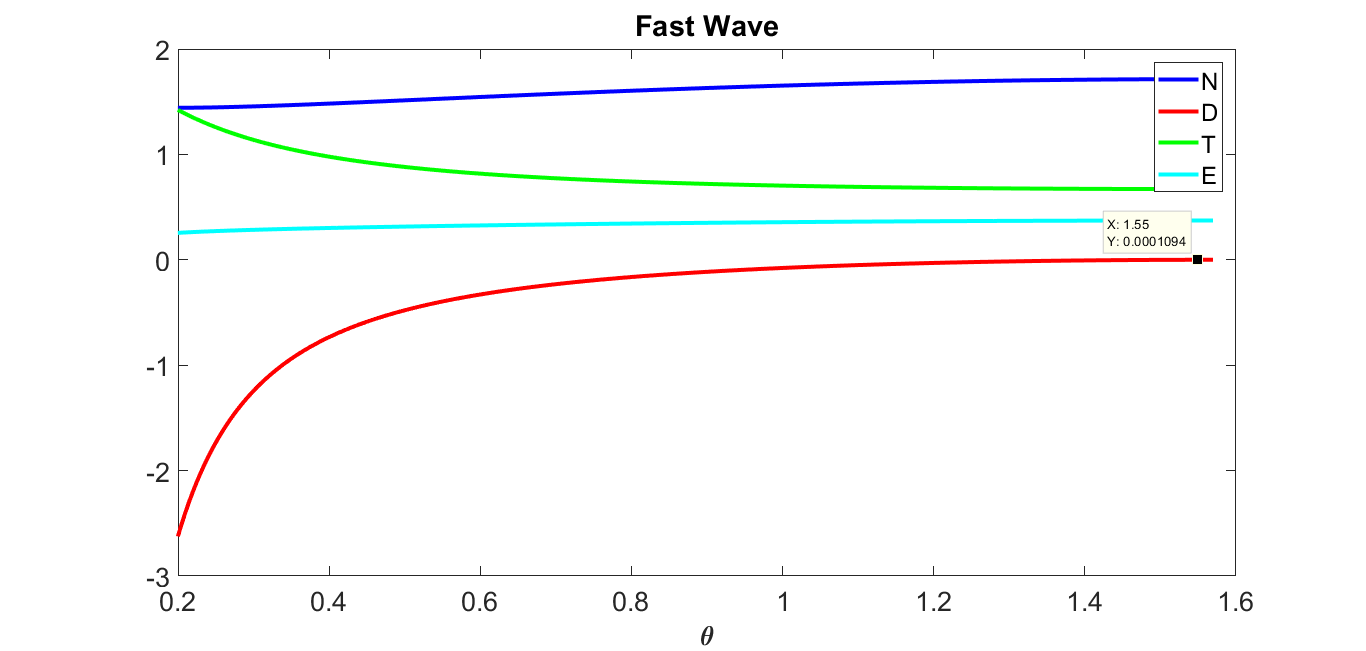}
\caption{Variations of coefficients against angle of propagation $\theta$ for $\beta=0.8$ and $\mu=1/1836$ for both slow and fast waves with $\theta$ expressed in radian unit}
\label{NDTEplot}
\end{figure}

At first sight, the nonlinear evolution equation (\ref{FinalKP}), which is derived in the previous section, seems to be a forced KP-II equation \cite{Yong}. For a detailed analysis of the nature of forced KP equation (\ref{FinalKP}), we plot the coefficients $N,\,D$, $T$ and $E$ against angle of propagation $\theta$ for constant $\beta$ which is as shown in Figure-\ref{NDTEplot}.
  From the figure, it is clear that $N$ and $D$ are positive whereas $T$ and $E$ are negative for slow magnetosonic waves. On the other hand, for fast magnetosonic waves, $N,\,T$ and $E$ are all positive whereas $D$ is negative when propagation angle $\theta < {\theta}_c$ and positive when $\theta > {\theta}_c$. Here it should be noted that ${\theta}_c$ is the angle at which the value of $D$ becomes zero; where our theory ceases to be valid. From the expression of $D$ in equation 
(\ref{NDTE}), the value of $\theta_c$ is calculated as
\begin{equation}
\theta_c =cos^{-1}(\sqrt{\frac{\mu}{1-\mu+{\mu}^2}}v_p) \label{ThetaC}.
\end{equation}
Therefore, as $v_p$ again depends upon $\beta$ and $\theta$ as per dispersion relation (\ref{DispRela}) derived in the previous section and $\mu$ is a constant given by equation (\ref{beta}), the value of ${\theta}_c$ is solely determined from the value of $\beta$; which depends upon $c_s$ and $v_A$ as per equation (\ref{beta}).

From the above analysis, it is clear that for the case of slow magnetosonic waves, the nonlinear evolution equation (\ref{FinalKP}) becomes forced KP-I equation as $T$ is negative. Similarly, for the case of fast magnetosonic wave, the nonlinear evolution equation (\ref{FinalKP}) becomes forced KP-I equation for propagation angle $\theta < \theta_c$ and forced KP-II equation when $\theta > \theta_c$. It should be noted in this context that when we neglect electron inertial effects, i.e. $\mu =0$, then $\theta_c=\pi /2$ as evident from equation (\ref{ThetaC}). This implies that for the entire region in parameter space corresponding to both slow and fast magnetosonic waves, we get forced KPI equation for this limiting case; which is in agreement with our recent work \cite{Acharyaa}, where electron inertial effects are neglected. For further analysis of both slow and fast magnetosonic waves, we follow the recent work done by Ruderman \cite{Ruderman} on stability of magnetosonic solitons in Hall plasmas. From his analysis of stability of magnetosonic solitons, it is concluded that magnetosonic solitons are stable with respect to transverse perturbations provided they are governed by a KP-II equation, and they are unstable with respect to transverse perturbations provided they are governed by a KP-I equation. For the case of KP-I equation, rational algebraic lump wave solution becomes stable. As we have obtained forced KP-I equation for slow magnetosonic wave and also for fast magnetosonic wave when $\theta< \theta_c$, lump solutions are stable in these cases. Only for fast magnetosonic waves, when $\theta>\theta_c$, magnetosonic solitary wave solutions are stable. These dynamical behaviours of both slow and fast magnetosonic waves are explored in detail in this section.


In recent years, Kumar and Sen \cite{Kumar} have observed excitations of magnetosonic solitary waves in presence of charged space debris particles in ionospheric plasma environment for the choice of ambient magnetic field to be in $Z$ direction. Most interestingly, curving nature of magnetosonic precursors is reported in their work when a $2D$ circular source is considered having perpendicular propagation with respect to ambient magnetic field. Their work is performed through particle in cell (PIC) simulation. In our recent work on space debris \cite{Acharya}, bending phenomena of electrostatic dust ion acoustic solitary waves is explained analytically; where we have derived a special exact curved solitary wave solution of the nonlinear evolution equation. However, we have not considered ambient magnetic field in the LEO region in this work. In our most recent work \cite{Acharyaa} on space debris, ambient magnetic field in the LEO region is considered analytically to determine debris dynamics in the framework of Hall MHD; where we have detected accelerated magnetosonic lump solutions accompanied with changes in their amplitudes. Therefore, $1D$ solitary waves generated due to space debris as reported in \cite{Sen, Truitt, Mukherjee} are liable to show lump like behaviours in two dimensions. In addition, there can also be possibilities of obtaining planar and ring solitons due to debris as reported in \cite{TruittKP, Acharya} that may be associated with special features like bending in spatial dimensions and time-dependent velocities showing accelerating behaviours. The bending phenomenon of magnetosonic solitary waves in $2D$ and $3D$ has hitherto not been investigated analytically in the context of space debris. In following subsections, these bending phenomena of magnetosonic waves are also discussed briefly. Therefore, this article can also corroborate the work of Kumar and Sen \cite{Kumar} who have concluded bending nature of magnetosonic solitons in spatial dimensions due to debris.

\subsection{Exact rational lump wave solutions}
As per our previous analysis in the parameter space of equation (\ref{FinalKP}), magnetosonic lump wave solutions are stable for slow waves and a large region in parameter space for fast waves where the condition for angle of propagation $\theta<\theta_c$ is satisfied. In order to explore the dynamics of these magnetosonic lump wave solutions in detail, we use the following redefinition of variables in equation (\ref{FinalKP}) for slow waves:
\begin{equation}
\tau =T;\,\xi={\lvert D \rvert}^{1/3}X;\,\zeta ={(\frac{{\lvert D \rvert}^{1/3}\lvert T \rvert}{3})}^{1/2}Z;\,n_1=\frac{6{\lvert D \rvert}^{1/3}}{N}U;\,J_D=\frac{6{\lvert D \rvert}^{1/3}}{EN}J, \label{NormKPLump}
\end{equation}
Then equation (\ref{FinalKP}) converges to
\begin{equation}
\frac{\partial}{\partial X}(\frac{\partial U}{\partial T}+6 U \frac{\partial U}{\partial X}+\frac{{\partial}^3 U}{\partial {X}^3})-3\frac{{\partial}^2 U}{\partial {Z}^2}=\frac{\partial J}{\partial {X}}. \label{FinalKPNormLump}
\end{equation}
It should be noted  that, we have to apply  another transformation, given as: $n_1\longrightarrow-n_1$; $\tau \longrightarrow -\tau$, to equation (\ref{FinalKP}) for getting the above equation (\ref{FinalKPNormLump}) in case of fast magnetosonic waves when $\theta<\theta_c$. This equation (\ref{FinalKPNormLump}) stands for the final normalized form of the nonlinear evolution equation for slow magnetosonic waves, and for fast magnetosonic waves when $\theta<\theta_c$. This normalized equation (\ref{FinalKPNormLump}) will be considered for further analysis in this subsection.

Recently, the work of Truitt et al. \cite{Truitt, TruittKP} reveals the fact that forcing debris function is derived to be of Gaussian nature. They have considered both $1D$ and $2D$ Gaussian functions representing effects of space debris particles in their work through simulations from forced KdV equation and forced KP equation respectively. According to their work, it depends on size, shape, velocity and surface potential of charged debris, Debye shielding effects due to surrounding plasma and distance of point of observation from debris. It can be easily seen that Gaussian function behaves like a $sech^2-$type structure in $1D$ and an algebraic lump wave type structure in $2D$. We know that $sech^2-$type structures are exponentially localized functions in certain directions whereas lump waves are completely localized functions which decrease in all directions. For some typical lump wave type localized forcing debris functions, we have investigated corresponding dynamical behaviours of analytical solution $U$ in our recent work \cite{Acharyaa} through exact and perturbative solutions of forced KPI equation.

Thus, for slow magnetosonic waves in the entire parameter space and fast magnetosonic waves in the parameter space where $\theta<\theta_c$, we can get exact pinned lump wave solution $U$ for the forced KP equation (\ref{FinalKPNormLump}) as:

\begin{equation}
U=4\frac{-{[X+\int A(T)dT+mZ+3(m^2-n^2)T]}^2+n^2{(Z+6mT)}^2+\frac{1}{n^2}}{{\lbrace {[X+\int A(T)dT+mZ+3(m^2-n^2)T]}^2+n^2{(Z+6mT)}^2+\frac{1}{n^2} \rbrace}^2}, \label{PinnedLump1}
\end{equation}
or in a compact notation
as 
\begin{equation}
 U = 4 \frac{[- \psi_1^2 + \psi_2^2 + 1/n^2]}{[\psi_1^2 + \psi_2^2 + 1/n^2]^2},\label{PinnedLump2}
\end{equation}

\begin{equation}
 \psi_1 = [{X}+ \int A(T)dT + m {Z}+3(m^2-n^2) {T} ], \ \psi_2 = n{[{Z}+6m {T}]}, 
\end{equation}
for the choice of debris function $J$ of (\ref{FinalKPNormLump}) as 
\begin{equation}
J = -8A(T)\frac{{\psi}_1 (-{\psi}^2_1+3{\psi}^2_2+\frac{3}{n^2})}{{({\psi}^2_1+{\psi}^2_2+\frac{1}{n^2})}^3}. \label{JExpanded}
\end{equation}
The two solutions (\ref{PinnedLump2}) and (\ref{JExpanded}) move together with the same velocity which can be understood by plotting $U$ and $J$ at different times as shown in \cite{Acharyaa}. Finding an exact solution is difficult for a forced nonlinear evolution equation because the forcing term generally destroys the exact solvability. But we have managed to find exact lump solutions for (\ref{FinalKPNormLump}) having a special property. Both $U$ and $J$ accelerate depending on the arbitrary function $A(T)$ which is the time dependent amplitude of the forcing function $J$.
It can be noted that at $\psi_1=\psi_2 =0$ the forcing function $J$ becomes zero whereas $U$ is maximum and equals to $4n^2$.

The maximum amplitude of the lump wave solution $U$ is attained at $\psi_1 = \psi_2 =0.$ Hence, the velocity of the maximum amplitude point of the wave can be determined as  
\begin{equation}
 \frac{d \psi_2}{dT} = 0 \longrightarrow V_{UZ} = \frac{dZ}{dT} = -6m.
\end{equation}
Similarly, we can get 
\begin{equation}
 \frac{dX}{dT} = V_{UX}=3(m^2+n^2)-A(T),
\end{equation}
where,
$
\vec{V}_U=(V_{UX},V_{UZ})$
 denote the velocity with which the accelerated lump wave solution (\ref{PinnedLump1}) moves, and $V_{UX}$ and $V_{UZ}$ represent the $X$ and $Z$ components of $\vec{V}_U$ respectively. Hence, the acceleration associated with this solution  (\ref{PinnedLump1}) can be evaluated as:
\begin{equation}
\vec{W}_U=(W_{UX},W_{UZ});\,W_{UX}=-\frac{dA(T)}{dT};\,W_{UZ}=0, \label{AccUaVel}
\end{equation}
where $\vec{W}_U$ denotes the acceleration of the solution (\ref{PinnedLump1}), and $W_{UX}$ and $W_{UZ}$ represent its $X$ and $Z$ components respectively. From the above equation (\ref{AccUaVel}), it is clear that acceleration of lump wave $U$ is coming due to presence of the term $A(T)$  in equation (\ref{PinnedLump1}).
The maximum amplitude of the accelerated lump wave solution (\ref{PinnedLump1}) is attained at $\psi_1=0=\psi_2$ and given as = $4 n^2$.

\begin{figure}[hbt!]
\centering
\includegraphics[width=10cm]{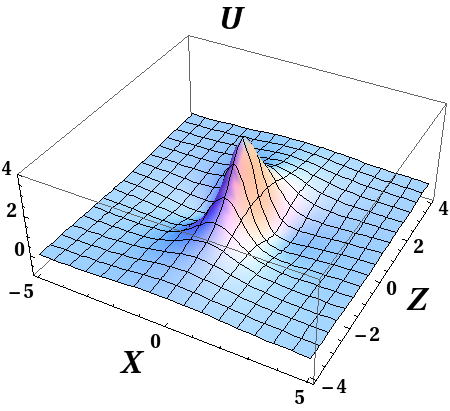}

\caption{Lump wave solution $U$ of equation (\ref{FinalKPNormLump}) in absence of external forcing, i.e. $A(T)=0$, on $X-Z$ plane for free parameters $m=0$ and $n=1$ at $T=0$. As usual, the lump wave decays algebraically in all directions} \label{DecayU} 
\end{figure}

\begin{figure}
\includegraphics[height=8cm]{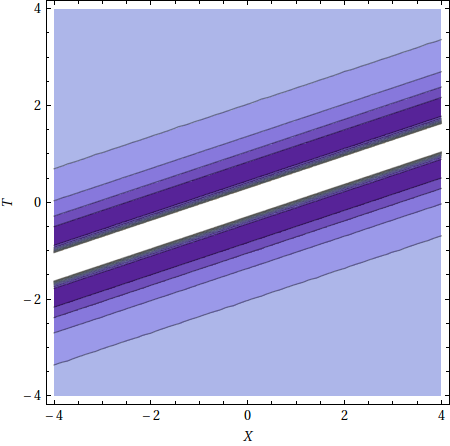}
 \ \ \ \ \includegraphics[height=8cm]{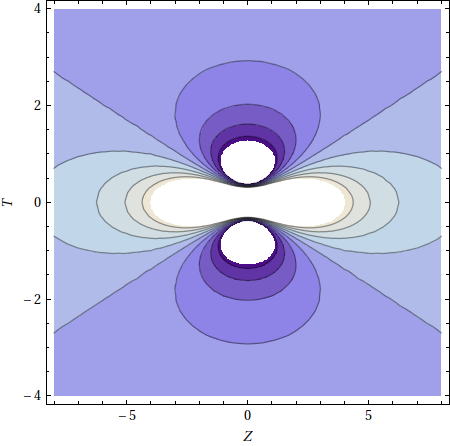}

\vspace{0.5cm}

\qquad \qquad 
(a) Contour plot of $U$ (\ref{PinnedLump1}) on $X-T$ plane at Z=0
\qquad \qquad
(b) Contour plot of $U$ (\ref{PinnedLump1}) on $Z-T$ plane at X=0.
\vspace{0.5cm}
\caption{The above figures are plotted for $A(T)=0$ which implies the absence of external forcing term $F$. Since we have chosen $m=0$ and $Z=0$ in sub-figure (a), the wave propagates along a straight line; as evident from equation (\ref{PinnedLump1}). Whereas in sub-figure (b), we have chosen $X=0$ and the wave does not propagate along $Z$ direction since $m=0$, hence we get a localized plot } \label{f4}
\end{figure}

\begin{figure}
\includegraphics[height=8cm]{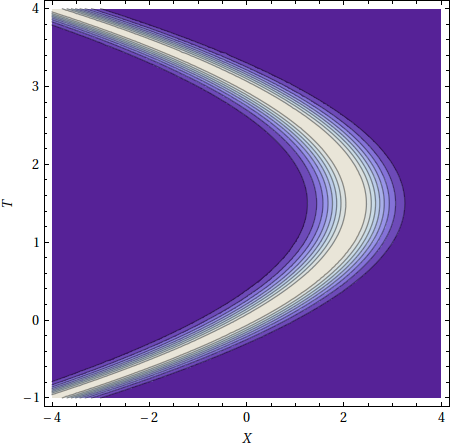}
 \ \ \ \ \includegraphics[height=8cm]{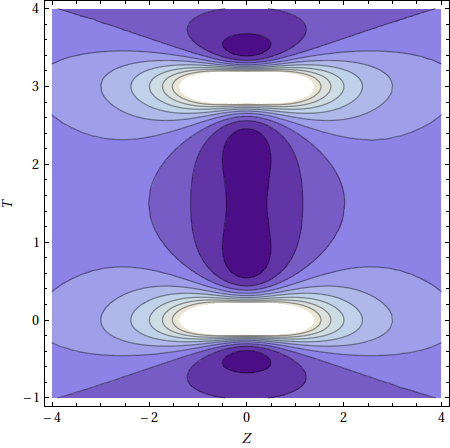}

\vspace{0.5cm}

\qquad \qquad 
(a) Contour plot of $U$ (\ref{PinnedLump1}) on $X-T$ plane at Z=0
\qquad \qquad
(b) Contour plot of $U$ (\ref{PinnedLump1}) on $Z-T$ plane at X=0
\vspace{0.5cm}
\caption{The above figures are plotted for $A(T)=2T$ with free parameters $m=0$ and $n=1$. It can be seen that, due to presence of forcing $F$ as $A(T)\neq0$, the solution bends in a parabolic path on $X-T$ plane as evident from the solution (\ref{PinnedLump1}). This is shown in sub-figure (a). Whereas in sub-figure (b), the localized plot gets distorted accordingly on $Z-T$ plane} \label{f5}
\end{figure}

\begin{figure}
\includegraphics[height=8cm]{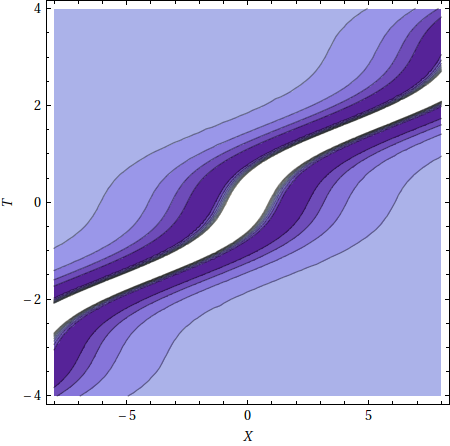}
 \ \ \ \ \includegraphics[height=8cm]{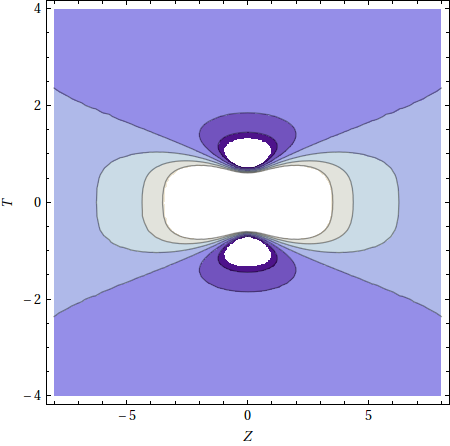}

\vspace{0.5cm}

\qquad \qquad 
(a) Contour plot of $U$ (\ref{PinnedLump1}) on $X-T$ plane at Z=0
\qquad \qquad
(b) Contour plot of $U$ (\ref{PinnedLump1}) in $Z-T$ plane at X=0
\vspace{0.5cm}
\caption{The above figures are plotted for a periodic choice of $A(T)$ as $A(T) = 2 \cos(2T)$ with free parameters $m=0$ and $n=1$. It can be seen that the solution bends on $X-T$ plane periodically as evident from the solution (\ref{PinnedLump1}). This is shown in sub-figure (a). Whereas in sub-figure (b), the localized plot gets distorted accordingly on $Z-T$ plane.}\label{f6}
\end{figure}

\begin{figure}
\includegraphics[height=8cm]{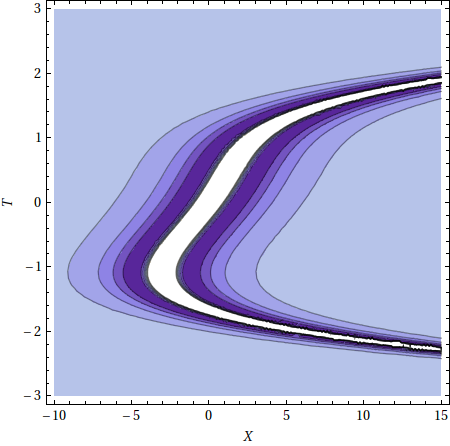}
 \ \ \ \ \includegraphics[height=8cm]{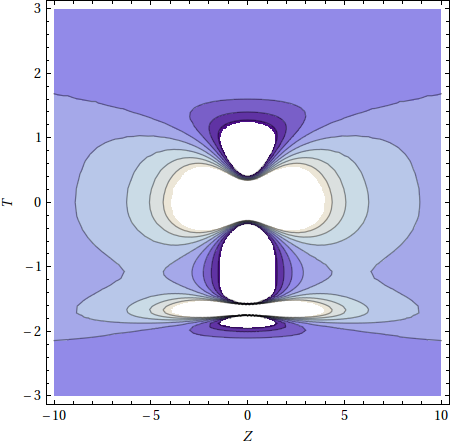}

\vspace{0.5cm}

\qquad \qquad 
(a) Contour plot of $U$ (\ref{PinnedLump1}) on $X-T$ plane at Z=0
\qquad \qquad
(b) Contour plot of $U$ (\ref{PinnedLump1}) on $Z-T$ plane at X=0
\vspace{0.5cm}
\caption{The above figures are plotted for $A(T) = 2 T - 4 T^3$ with free parameters $m=0$ and $n=1$. It can be seen that the solution bends on $X-T$ plane like a double well potential as evident from the solution (\ref{PinnedLump1}). This is shown in sub-figure (a). Whereas in sub-figure (b), the localized plot gets distorted accordingly on $Z-T$ plane}\label{f7}
\end{figure}

Similarly, the velocity and acceleration of the null point, i.e. where $J=0$, of the forcing function can
be calculated. It turns out that the velocity and acceleration of the point $\psi_1=\psi_2 = 0$ of the forcing function $J$, is the same as that of $U$. Hence they can be called as pinned accelerated lump wave solutions because the propagate together. But we can change the velocity of $U$ and $J$ accordingly by the choice of the arbitrary function of time $A(T)$. Thus we can mathematically model the charged debris induced magnetosonic lump waves for the real physical condition by appropriate choices of $A(T)$. For $A(T)= constant$,  we would get the standard lump solutions of KP-I equation travelling with constant velocity. We have plotted the solution $U$ in absence of external forcing for $A(T) = 0$ in Figure-\ref{DecayU} and Figure-\ref{f4}. The detailed dynamics is further described through contour plots in Figure-\ref{f5}, Figure-\ref{f6} and Figure-\ref{f7} for different choices of $A(T)$ to represent effects of space debris. Since the wave gets accelerated, the change in the shape of the contour plots on $X-T$ and $Z-T$ planes can clearly describe the dynamics.

Actually, in this case, the forced KP-I equation can be converted to unforced KP-I equation by a frame transformation. Similarly, we can generate also the higher order pinned accelerated lump wave solutions of (\ref{FinalKPNormLump}) in an analogous manner.

\subsection{Curved solitary wave solutions}
As per our previous analysis in the parameter space of equation (\ref{FinalKP}), magnetosonic solitary waves are stable for a small region in parameter space for fast waves where the condition for angle of propagation $\theta>\theta_c$ is satisfied. In order to further explore dynamics of these magnetosonic solitary waves in detail, we use the same redefinition of variables as given by equation (\ref{NormKPLump}) in the previous subsection in nonlinear evolution equation (\ref{FinalKP}) to obtain

\begin{equation}
\frac{\partial}{\partial X}(\frac{\partial U}{\partial T}+6 U \frac{\partial U}{\partial X}+\frac{{\partial}^3 U}{\partial {X}^3})+3\frac{{\partial}^2 U}{\partial {Z}^2}=\frac{\partial J}{\partial {X}}, \label{FinalKPNormPlanar}
\end{equation}
 which is structurally similar to equation (\ref{FinalKPNormLump}) except for a negative sign in the second term on the left hand side. This equation (\ref{FinalKPNormPlanar}) represents the final normalized form of the nonlinear evolution equation for fast magnetosonic waves when $\theta>\theta_c$ in our system; which will be considered for further analysis in this subsection. 

 We have derived a forced KP-II equation (\ref{FinalKPNormPlanar}) as the evolution equation for nonlinear fast magnetosonic waves when $\theta>\theta_c$ in presence of charged space debris particles. As discussed earlier, forcing function due to space debris is of Gaussian nature as reported in \cite{Truitt, TruittKP}. However, Sen et al. \cite{Sen} have discussed two exact line solitary wave solutions for their forced KdV equation in $(1+1)$ dimensions that have constant amplitude and velocity by using  localized forcing debris functions of $sech^2$ and $sech^4$ types. There have also been reported a few studies made in (1+1) dimensions by considering periodic forms of the debris function \cite{TFAli,TFChatterjee,TFZhen}.
Considering these facts, it can be seen that there can be several types of localized functions to be chosen as forcing debris functions; which need not be exactly Gaussian, but can represent similar behaviours. In our earlier work \cite{Acharya}, we have reported both accelerated and curved dust ion acoustic solitary wave solutions for some particular localized forcing debris functions with nonlinear evolution equation as forced KP-II equation. Therefore, for these forcing debris functions, we can get accelerated and curved magnetosonic solitary wave solutions when $\theta > \theta_c$ for fast waves following the same procedure as reported in \cite{Acharya}. Moreover, different analytic solitary wave solutions can be obtained for forced KP equation following \cite{Zhao}.

\begin{figure}
\includegraphics[height=8cm]{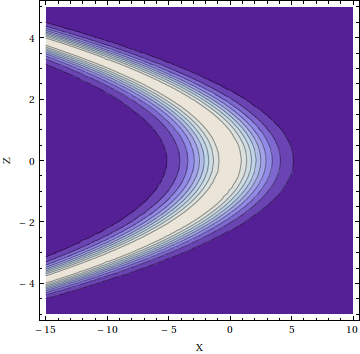}
 \ \ \ \ \includegraphics[height=8cm]{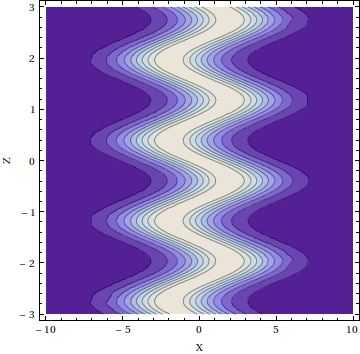}

\vspace{0.5cm}

\qquad 
(a) Contour plot of $U$ (\ref{csol}) on $X-Z$ plane for $A(Z) = Z^2$
(b) Contour plot of $U$ (\ref{csol}) on $X-Z$ plane for $A(Z) = 2\sin{(4Z)}$
\vspace{0.5cm}
\caption{The above figures are plotted for $c_1 = 0.5$ and $ \theta_0 = 0$ at $T=0$. It can be seen that the curved solitary wave  solution (\ref{csol})  bends on $X-Z$ plane for various choices of the arbitrary function $A(Z)$} \label{f8}
\end{figure}

\begin{figure}
\includegraphics[height=8cm]{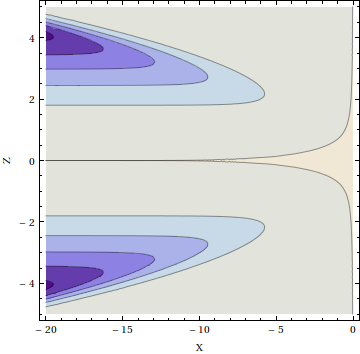}
 \ \ \ \ \includegraphics[height=8cm]{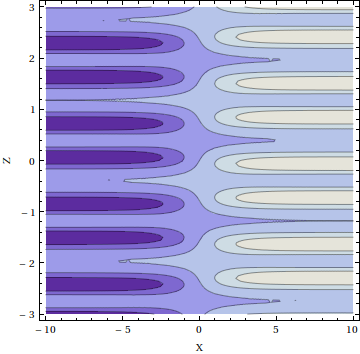}

\vspace{0.5cm}

\qquad 
(a) Contour plot of $J$ (\ref{Fcurv}) on $X-Z$ plane for $A(Z) = Z^2$
(b) Contour plot of $J$ (\ref{Fcurv}) on $X-Z$ plane for $A(Z) = 2\sin{(4Z)}$
\vspace{0.5cm}
\caption{The above figures are plotted for $c_1 = 0.5$ and $ \theta_0 = 0$ at $T=0$. It can be seen that debris function (\ref{Fcurv}) gets distorted on $X-Z$ plane for various choices of the arbitrary function $A(Z)$} \label{f9}
\end{figure}

We can derive  an exact pinned solitary wave solution $U$ of equation (\ref{FinalKPNormPlanar}), for a special choice of forcing function $J$, that can bend on $X-Z$ plane as
\begin{equation}
 U = \frac{c_1}{2} \ Sech^2[\frac{\sqrt{c_1}}{2} \ (X + A(Z) - c_1 T + \theta_0)],\label{csol}
\end{equation} 
depending on the function $A(Z)$. The forcing function $J$ of (\ref{FinalKPNormPlanar})  can be simplified as 
\begin{align}
 &{} J = 3 A_{ZZ} U + 3 A_Z^2 \ U_{X}. 
 \label{Fcurv}
\end{align}
Here, $c_1$ is a constant and $A(Z)$ is an arbitrary  function of $Z$. The function $J$ in equation (\ref{Fcurv}) vanishes as $X \longrightarrow \pm \infty$. There might be other choices of $J$ to get exact curved solitary wave solution $U$ of (\ref{FinalKPNormPlanar}) but we consider (\ref{Fcurv}) as the first example of $F$ to explore the situation.
The nonlinear  function $A(Z)$ makes the solitary wave to bend on $X-Z$ plane. Both the solutions  $U$ and $J$ given by equations (\ref{csol}) and (\ref{Fcurv}) are pinned solutions. 
Thus, we have provided an exact pinned curved solitary wave solution $U$ for equation (\ref{FinalKPNormPlanar}) for a specific choice of the localized function $J$. The solution may be used for modelling the twist or turn of the magnetosonic solitary waves during its motion in real experiments. The contour plots of $U$ and $J$ for a few choices of $A(Z)$ are shown in Figure-\ref{f8} and Figure-\ref{f9}.

Therefore, the gist of this section lies in the fact that we have obtained stable magnetosonic lump wave solutions for slow waves in the entire parameter space and fast waves in a large region of parameter space where $\theta<\theta_c$ condition is satisfied. In contrast, in a small region of parameter space corresponding to fast waves, where $\theta>\theta_c$ is valid, magnetosonic curved solitary wave solutions are reported to be stable. As value of $\theta_c$ normally lies closer to $\pi / 2$, only for angles of propagation nearer to $\pi / 2$, magnetosonic curved solitary wave solutions are found to be stable whereas magnetosonic lump wave solutions become unstable. It can be seen, in comparison with our earlier work \cite{Acharyaa}, that inertial effects are important only for angles of propagation nearer to $\pi /2$ in case of fast waves; where instability of magnetosonic lump solutions is reported and magnetosonic curved solitary wave solutions are obtained as stable solutions. Apart from this special case, inertial effects are not that significant because they cannot affect the stability of magnetosonic lump wave solutions, i.e. whether these effects are considered or not in the formulation does not effectively alter the dynamics of magnetosonic waves.
 The bending phenomena of fast magnetosonic solitary waves, as discussed earlier for two-dimensional sources, for nearly perpendicular propagations corroborate work of Kumar and Sen \cite{Kumar}; who have concluded non-uniform spatial dependences or curving nature of fast magnetosonic solitary waves due to a two-dimensional circular source term. Also, their work is performed for perpendicular propagation only. 

 In our work, magnetosonic lump solutions have widths of the order of ion inertial lengths except for one exception as following. For nearly parallel propagations, magnetosonic lump solutions have widths of the order of Debye lengths as reported in \cite{Kennel}. Similarly, for nearly perpendicular propagations, finite electron inertial effects change the sign of dispersion through angle $\theta_c$ as discussed earlier and thereby reducing the scale length to order of electron skin depths or electron inertial lengths in case of curved fast magnetosonic solitary waves.
 This work is also in perfect agreement with \cite{Kumar} by reproducing curved fast magnetosonic solitary waves for nearly perpendicular propagations having exactly the same widths as that in \cite{Kumar}; thereby permitting large footprints in ionospheric plasma conditions in presence of space debris that enables their easy detection. In this context, it is also important that magnetosonic lump wave solutions reported in our earlier work on space debris in the framework of Hall plasma have widths of the order of ion inertial lengths for both slow and fast waves that is very much favourable for their easy detection. In addition, bending phenomena or distortion of lump waves in spatial dimensions is also a possibility, and is applicable to a wider region of parameter space. This enables study of space debris excited nonlinear structures in a wider context.
 

\section{Conclusive Remarks}
Conclusively we state that we have derived a forced Kadomtsev-Petviashvili  equation in this work describing the nonlinear evolution of magnetosonic waves in presence of charged space debris in the low Earth orbital plasma region using inertial MHD. For several choices of space debris function, stable accelerated and curved solitary wave solutions of the forced KP equation are obtained in different regions of parameter space for both slow and fast magnetosonic waves. These solitary wave solutions remain pinned to forcing debris functions. In particular, for nearly perpendicular propagations, retaining the effects of electron inertia leads to curved solitary wave solutions for fast magnetosonic waves. These solitary waves are again pinned in nature and travel with the same velocity as source debris while also exhibiting a curved nature. This is the most crucial finding of this article.

 \section{Acknowledgements}
 Siba Prasad Acharya acknowledges the financial support received from Department of Atomic Energy (DAE) of Government of India during this work through institute fellowship scheme. Abhik Mukherjee acknowledges Indian Statistical Institute, Kolkata, India for the financial support during the progress of the work.
 

\section{References} \label{Refer}
\begin{thebibliography}{99}
\bibitem{Sen} A. Sen, S. Tiwari, S. Mishra, and P. Kaw, Advances in Space Research, 56 (2015) 3 429-435.
\bibitem{Mukherjee} A. Mukherjee, S. P. Acharya, and M. S. Janaki, Astrophys Space Sci, 366 (2021) 7.
\bibitem{Acharya} S. P. Acharya, A. Mukherjee, and M. S. Janaki arXiv:2010.06901v2 [physics.plasm-ph].
\bibitem{Acharyaa} S. P. Acharya, A. Mukherjee, and M. S. Janaki arXiv:2103.06593 [physics.plasm-ph].
\bibitem{Kulikov} I. Kulikov and M. Zak,
Advances in Remote Sensing, 1 (2012) 3.
\bibitem{Truitt} Alexis S. Truitt, and Christine M. Hartzell, Journal of Spacecrafts and Rockets, 57 (Published online on 24 Apr 2020) 5.
\bibitem{TruittKP} Alexis S. Truitt, and Christine M. Hartzell, Journal of Spacecrafts and Rockets, Article in Advance (Published online on 1 Dec 2020).
\bibitem{Sen3} G. Arora, P. Bandyopadhyay, M. G. Hariprasad and A. Sen, Physics of Plasmas, 26 (2019) 093701.
\bibitem{Kumar} A. Kumar, and A. Sen, New J. Phys., 22 (2020) 073057.
\bibitem{Ruderman} M. S. Ruderman, Physica Scripta, 95 (2020) 095601.
\bibitem{Sampaio} J. C. Sampaio, E. Wnuk, R. Vilhena de Moraes, and S. S. Fernandes, Mathematical Problems in Engineering, Article ID 929810 (2014).

\bibitem{Lingam} M. Lingam, P. J. Morrison, and E. Tassi, Physics Letters A, 379 (2015) 570–576.
\bibitem{Janaki} M. S. Janaki, B. K. Som, B. Dasgupta, and M. R. Gupta, Journal of the Physical Society of Japan, 60 (1991) 9 2977-2984.
\bibitem{Yong} X. Yong, W. X. Ma, Y. Huang, and Y. Liu, Computers and Mathematics with Applications 75 (2018) 3414–3419.










\bibitem{TFAli} R. Ali, A. Saha and P. Chatterjee, Physics of Plasmas,  24 (2017)  122106.

\bibitem{TFChatterjee} P. Chatterjee, R. Ali and A. Saha, Zeitschrift fur Naturforschung A,  73 (2018) 861.

\bibitem{TFZhen} H. Zhen, B. Tian, H. Zhong, W. Sun and M. Li, Physics of Plasmas,  20  (2013) 082311.

\bibitem{Zhao} Z. Jun-Xiao, and G. Bo-Ling, Commun. Theor. Phys., 52 (2009) 279–283.
\bibitem{Kennel} C. F. Kennel, B. Buti, T. Hada, and R. Pellat, Physics of Fluids, 31 (1988) 1949.


\end{thebibliography}
\end{document}